\documentclass[aps,prb,print,superscriptaddress,nofootinbib,longbibliography,floatfix,showkeys]{revtex4-2}
\usepackage[utf8]{inputenc}
\usepackage[T1]{fontenc}

\usepackage{amsfonts}
\usepackage{amsmath}
\usepackage{amssymb}
\usepackage{graphicx}
\usepackage{xcolor}
\usepackage{tabularx}
\usepackage{booktabs}
\usepackage{subfigure}
\usepackage{bm}

\usepackage[italicdiff]{physics}
\usepackage[protrusion=true,expansion=true]{microtype}
\usepackage[colorlinks=True,allcolors=blue]{hyperref}
\usepackage{cleveref}
\usepackage{orcidlink}
\allowdisplaybreaks

\newcommand{\mc}[1]{\mathcal{#1}}

\newcommand{\bS}{\bm{S}}
\newcommand{\onehalf}{\frac{1}{2}}
\newcommand{\threehalf}{\bm{\frac{3}{2}}}

\newcommand{\UGxIFTiA}{Institute of Theoretical Physics and Astrophysics, Faculty of Mathematics, Physics and Informatics, University of Gda\'nsk, ul. Wita Stwosza 57, 80-308 Gda\'nsk, Poland}

\newcommand{\UGxICTQT}{International Centre for Theory of Quantum Technologies, University of Gda\'nsk, ul. Ba\.zy\'nskiego 9, 80-309 Gda\'nsk, Poland}

\begin{document}

\title{Micromagnons and long-range entanglement in ferrimagnetic ground states}

\author{Marcin Wie{\'s}niak\,\orcidlink{0000-0002-1873-3945}}
\email{marcin.wiesniak@ug.edu.pl}
\affiliation{\UGxIFTiA}

\author{Ankit Kumar\,\orcidlink{0000-0003-3639-6468}}
\affiliation{\UGxIFTiA}

\author{Idriss Hank Nkouatchoua Ngueya\,\orcidlink{0000-0003-4779-2677}}
\affiliation{\UGxICTQT}
\affiliation{\UGxIFTiA}

\begin{abstract}
While significant attention has been devoted to studying entanglement in photonic systems, solid-state spin lattices remain relatively underexplored. Motivated by this gap, we investigate the entanglement structure of one-dimensional ferrimagnetic chains composed of alternating spin-1/2 and spin-3/2 particles. We characterize the ground-state correlations using exact diagonalization and the Density Matrix Renormalization Group method. Although the bipartite entanglement is restricted to nearest neighbors, we reveal the presence of long-range genuine multipartite entanglement between spatially separated spin pairs. These findings advance our understanding of quantum correlations in ferrimagnetic materials. The micromagnon description allows to provide fast approximation of ground states of ferrimagnets and emphasizes presence of multipartite correlations not widely discussed thus far.
\end{abstract}

\maketitle

\section{Introduction}

Quantum entanglement refers to correlations between subsystems that surpass what is achievable through coordinated local operations alone \cite{schrodinger1935gegenwartige}. Entanglement is commonly regarded as a feature of relatively small quantum systems, and it has a high fragility to environmental disturbances, such as interactions with a heat bath, particle loss, or other experimental limitations. Nevertheless, in many solid-state models, entanglement arises naturally in low-temperature regimes due to exchange interactions between constituent spins.
The landmark Bell’s theorem \cite{bell1964einstein} sparked tremendous interest in entanglement, not only from the standpoint of quantum foundations but also as a crucial resource in quantum technologies. Entanglement underpins protocols in quantum cryptography \cite{Ekert_QuantCrypto}, quantum computation \cite{jozsa2003role}, etc. 
In quantum communication, the goal is to utilize on-demand, fast-moving entanglement carriers, making photons ideally suited for such applications. In contrast, quantum computation favors stationary and controllable systems, attributes naturally present in bulk spin lattices.

Another reason for a growing interest in entanglement in solid-state systems is the recognition that many macroscopic properties of materials depend critically on their underlying quantum structure. While phenomena like crystal shape or color already necessitate a quantum description, certain low-temperature thermodynamic properties - such as free energy~\cite{dowling2004energy}, heat capacity~\cite{wiesniak2008heat}, and magnetic susceptibility~\cite{wiesniak2005magnetic}, can only be explained accurately by considering quantum correlations among individual spins.
Among the various physical models, two classes of materials stand out for their relevance to entanglement studies. The first comprises antiferromagnets, which are lattices of identical spins coupled via the Heisenberg interaction. In such systems, quantum correlations are often inferred from anomalously low magnetic susceptibility values \cite{wiesniak2005magnetic}.
The second class comprises ferrimagnets, which consist of sublattices with differing spin magnitudes, coupled like antiferromagnets. Although the sublattices tend to cancel each other's magnetization, one typically dominates. This leads to both the spontaneous and the field-induced magnetization, albeit weaker than in ferromagnets, where spins align uniformly. While this may obscure the detection of entanglement via magnetic susceptibility, the energetic preference for antiparallel neighboring spins suggests the presence of strong quantum correlations.

Despite their widespread practical applications in everyday devices, ferrimagnets are yet to be thoroughly analyzed at the microscopic level. The first theoretical framework for ferrimagnetism (along with antiferromagnetism) was proposed by N\'eel \cite{neel1948proprietes}, long before entanglement emerged as a major focus in quantum theory. Contemporarily, ferrimagnets are often studied in context of their usability in sensing and data storage. 
Specifically, effects such as ultrafast optical magnetization  switching~\cite{PhysRevLett.99.047601} and colossal magnetoresistance~\cite{li2015giant,PhysRevB.103.L161105,PhysRevB.106.L180402}. 
These systems may also display phase transitions \cite{tenorio2011quantum,PhysRevB.96.214419}. Substantial research has since investigated quantum correlations involving magnons (spin waves), phonons, and photons~\cite{wu2021remote,shim2022enhanced,qian2022entangling,fan2023entangling}. 
Studies focusing on spin entanglement in ferrimagnets include Refs. \cite{li2006entanglement,hao2007entanglement,solano2011entanglement,lee2025intersublattice}; notably, Ref.~\cite{vargova2024evidence} investigates a specific form of multi-site entanglement within a small ferrimagnetic cluster. Nevertheless, works such as Refs.~\cite{PhysRevB.95.174440,PhysRevB.106.L180402,susilo2024high,gu2024unconventional} continue to emphasize the need for a deeper understanding of the microscopic structure of ferrimagnets.

Several computational techniques exist to probe the ground-state features of spin lattices like the ones we investigate. Exact diagonalization offers highly accurate results but quickly becomes computationally prohibitive as system size increases. Nonetheless, the systems that are accessible using this method already provide reasonable approximations of the thermodynamic limit. Another technique widely employed in this study—particularly effective for one-dimensional systems—is the Density Matrix Renormalization Group (DMRG)~\cite{white1992density,ITensor}. However, by design, DMRG may distort long-range correlations and becomes less suitable for analyzing more complex lattice geometries. The Bethe Ansatz \cite{bethe1931theorie} also offers an exact solution for specific integrable models in terms of interacting spin waves, although it often comes with considerable computational overhead. Similarly, when applied directly,  the variational method can quickly exhaust the computational resources of even modern computers. Other approaches, such as quantum Monte Carlo simulations  \cite{ceperley1986quantum} and mean-field algorithms, often overlook correlations over longer distances.

Previous studies have reported the presence of long-range and genuine multi-site entanglement in various spin systems \cite{campos2006long,campos2007long,sahling2015experimental,consiglio2025long}. 
In ferrimagnetic chains, say, of spins-$\onehalf$ and $\threehalf$ (denoted as a spin-$\left(\onehalf,\threehalf\right)$ chain, bold font used for larger spins), the quantum correlations primarily arise from coherence between neighboring-site states such as $\ket{-\onehalf,\threehalf}$, and $\ket{\onehalf,\bm{\onehalf}}$.
However, the bipartite entanglement in these systems is restricted only to nearest neighbors, and hence we focus on our investigation of genuine multipartite across two adjacent spin pairs.
To build an intuitive understanding of entanglement properties in ferrimagnetic spin chains, we introduce the concept of a micromagnon ($\mu$-magnon)—an elementary excitation from the Néel state, as discussed earlier. We begin by analyzing the behavior of $\mu$-magnons, including their splitting and overlapping dynamics. We then study the relationships between amplitudes of various spin configurations. This analysis enables us to propose a computationally efficient scheme for approximating ground states in longer spin chains. While this method involves some precision trade-offs, its predictions align well with results obtained via DMRG. Importantly, this framework confirms the existence of long-range genuine multipartite entanglement.

\begin{figure*}
    \centering

\subfigure[]
{ \includegraphics[width=0.40\linewidth]{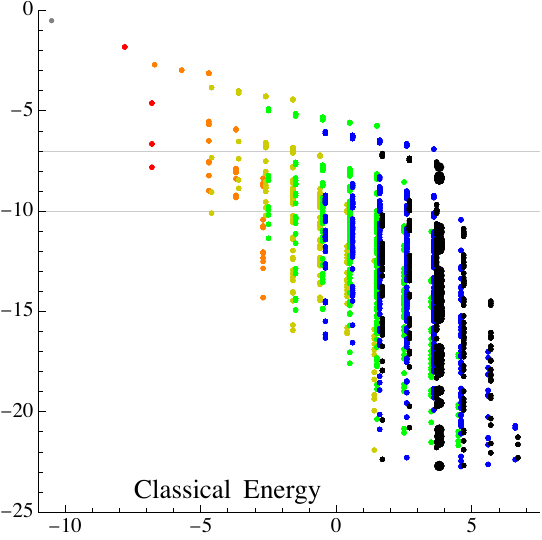} }
\hfill
\subfigure[]
{ \includegraphics[width=0.40\linewidth]{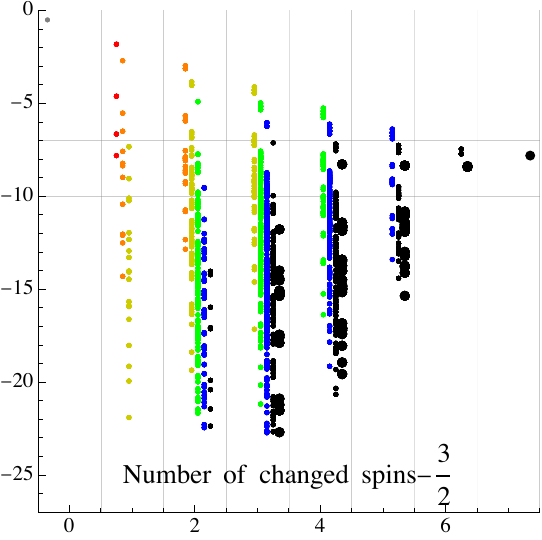} }
\hfill
\subfigure[]
{ \includegraphics[width=0.05\linewidth]{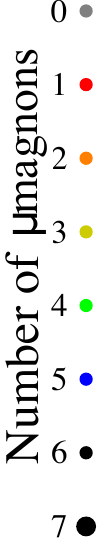} }

\caption{Logarithms of moduli $\log|\alpha|$ of amplitudes of configurations in the ground state of 14-site mixed spin chain versus the classical energy of the configurations (left) and the number of spin-$\threehalf$ changed with respect to the N\'eel sea (right). The coloring describes the number of $\mu$-magnons, i.e, the number of spins-$\onehalf$ changed with respect to the N\'eel sea.}
    \label{Amplitudes}
\end{figure*}

\section{Properties of ferrimagnetic ground-state}

We begin by investigating a linear system with alternating spin magnitudes, where the odd sites are occupied by spin-$\onehalf$ and the even sites by $\threehalf$, which was the first configurations to be considered by us. However, to highlight a more general behavior of the model, we also include few results for other spin combinations, mainly for spin-($\onehalf,\bm{1}$) systems.
The neighboring spins are coupled through the Heisenberg interaction with a constant strength $J$:
\begin{align}
H =
\sum_{k \in \text{odd}} J \,
\vec{S}^{\,[k]} \cdot \vec{\bS}^{\,[k+1]} - B \, S_z^{\,[k]}
+  \sum_{k \in \text{even}} J \,
\vec{\bS}^{\,[k]} \cdot \vec{S}^{\,[k+1]} - B \, \bS_z^{\,[k]} ,
\label{HAMIL}
\end{align}
where we have additionally applied a weak external magnetic field $B$ to remove the degeneracy of the ground state. To better understand an infinite chain, we primarily focus on the case of periodic boundary conditions, but eventually we take open ends into account and find a small correction. 
We also include results obtained with DMRG, which are in an excellent agreement with exact diagonalization values for shorter systems.
The transition between the chain and the ring is done by adding or removing to or from them Hamiltonian:
\begin{align}
    H' = J \,
\vec{\bS}^{\,[N]} \cdot \vec{S}^{\,[1]}  .
\end{align}
It may be convenient at times to rewrite \Cref{HAMIL} in terms of the raising and lowering operators:
\begin{align}
H =& \sum_{k \in \text{odd}} J \,
S^{\,[k]}_z \cdot \bS^{\,[k+1]}_z - B \, S_z^{\,[k]}
+ \sum_{k \in \text{even}} J \,
\bS^{\,[k]}_z \cdot S^{\,[k+1]}_z - B \, \bS_z^{\,[k]}
\nonumber\\[0.5em]
&+ \frac{J}{2} \sum_{k \in \text{odd}} S^{\,[k]}_+ \cdot \bS^{\,[k+1]}_- 
+ S^{\,[k]}_- \cdot \bS^{\,[k+1]}_+ 
+ \frac{J}{2} \sum_{k \in \text{even}}
  \bS^{\,[k]}_+ \cdot S^{\,[k+1]}_- 
+ \bS^{\,[k]}_- \cdot S^{\,[k+1]}_+ ,
\\[1em]
H' =&
J \,
\bS^{\,[N]}_z \cdot S^{\,[1]}_z + \frac{J}{2} \qty(
  \bS^{\,[N]}_+ \cdot S^{\,[1]}_- 
+ \bS^{\,[N]}_- \cdot S^{\,[1]}_+ ) .
\end{align}

The rotational symmetry at $B=0$ and the commutation between the Hamiltonian and the magnetic term allow us to express the Hamiltonian as block-diagonal in the sectors of fixed angular momenta as we explain below. 
For ferrimagnetic $N$-site spin-$(s_1,\mathbf{s}_2)$ lattices of alternating spins of magnitudes $s_1$ and $\bm{s}_2 > s_1$, the ground state is nested in one subspace with total angular momentum $N(\bm{s}_2-s_1)/2$. 
Applying a weak external magnetic field distinguishes the unique ground state with specific magnetization in the $z$ direction -- the one with magnetization $N(\bm{s}_2-s_1)/2$ in the direction of the field. 
Stronger magnetic fields can overcome the Heisenberg interaction and enforce a ground state with strong magnetization. 
All spin configurations belonging to these subspaces of a fixed magnetization differ from one another by an even number of states of individual sites. For instance, lowering the spin projection at a spin-$\threehalf$ site must be accompanied by raising the spin at a spin-$\onehalf$ site to conserve the total magnetization. This constraint structures the allowed transitions within the subspace.

The leading component of the ground state with  is the N\'eel state \cite{neel1948proprietes},
\begin{align}
\ket{\text{N\'eel}} = \underbrace{ \ket{ -{\onehalf}, +\bm{\threehalf}, \dots,  -{\onehalf}, +\bm{\threehalf} ,\dots, -{\onehalf}, +\bm{\threehalf} }}_{N \ \text{spins}}.
\end{align}
The amplitudes of the N\'eel state, i.e., its overlap with the true ground state $\ket{\Psi_0}$, for weak external magnetic fields can be approximated by
\begin{align}
   \alpha(\text{N\'eel}) =
  \braket{\text{N\'eel}}{\Psi_{0}}
  \approx -(-1)^{N/2} \, 0.99053 \times 0.96515^N .
\end{align}
where $N$ is the total number of sites.

Similar to the case of a spin-$\onehalf$ antiferromagnetic chain, the N\'eel state serves as a crude approximation of the actual ground state, as it only minimizes the energy among product states and does not account for quantum correlations. Consequently, for any classical configuration, regions that follow the N\'eel pattern can be referred to as ``the N\'eel sea'', while deviations from this pattern will be viewed as ``land''.
A basic structure within this framework is one that can be superimposed on the N\'eel sea, maintaining the total magnetization and avoiding intersections with other structures. Operationally, a structure is defined as a region where the magnetization, accumulated from a specific site, deviates from the N\'eel configuration.

\begin{figure}
    \centering

\subfigure[]
{ \includegraphics[width=0.66\linewidth]{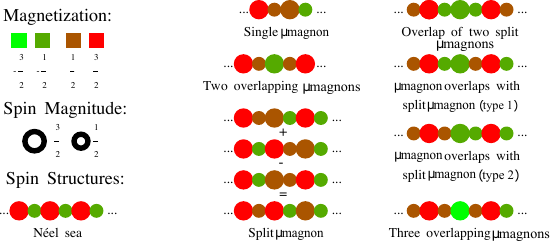} }
\hfill
\subfigure[]
{ \includegraphics[width=0.3\linewidth]{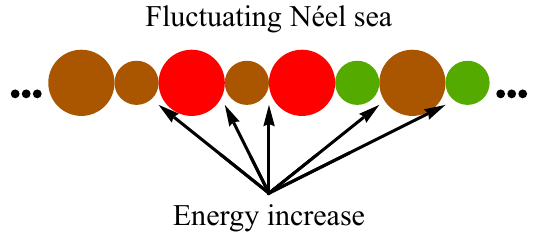} }

\caption{
(a)
Most typical spin structures identifiable in the ground state. Split $\mu$-magnons with larger separations are also possible. Magnetization $-\threehalf$ is attained when three $\mu$-magnons, at least one of which is split, overlap, which is associated with a very low amplitude.
(b)
A graphical representation of a classical mechanism of a $\mu$-magnon creation. The N\'eel sea, not being a true ground-state, is a subject to fluctuations. Distortions from the N\'eel pattern require less additional energy, when they are close together. Quantum-mechanical energy minimization dictates the amplitudes between fluctuating and non-fluctuating components.
}
\label{structure}
\end{figure}

The first hypothesis suggests that configurations with higher classical energy—calculated as the sum of the products of the $z$-components of neighboring spins while maintaining the correct total magnetization—should contribute less to the ground state. However, \Cref{Amplitudes} demonstrates this is not necessarily true. While we observe a slow exponential decay in the contribution of higher-energy configurations, components with intermediate energies exhibit a wide distribution, implying a more efficient ordering of components.
This ordering can be identified by sorting configurations according to the absolute value of their amplitudes, allowing us to recognize specific structures within the spin configuration. \Cref{structure}(a) presents the concept of N\'eel sea and examples of the most relevant spin structures, 
and in \Cref{config} in Appendix we additionally show a pictorial representation for some of the most pertinent configurations.

\begin{figure}
    \centering

\subfigure[Alternating spins $\onehalf-\threehalf$.]
{ \includegraphics[width=0.66\linewidth]{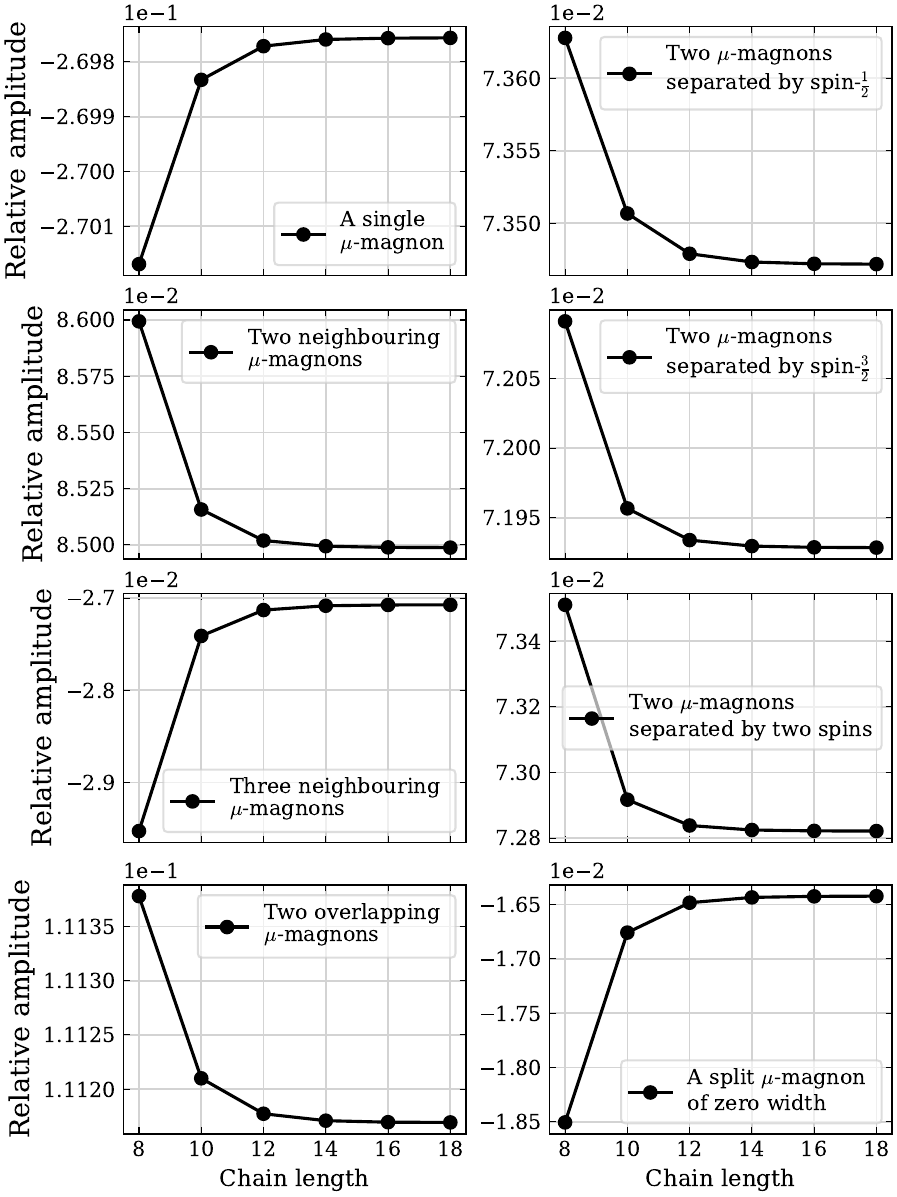} }

\subfigure[Alternating spins $\onehalf-\bm{1}$.]
{ \includegraphics[width=0.66\linewidth]{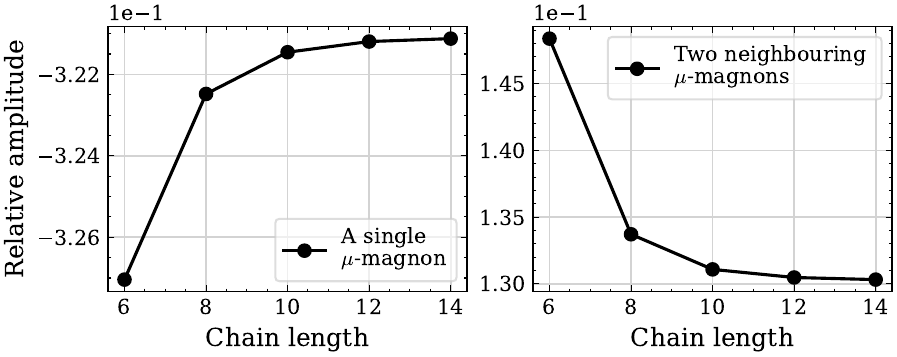} }

\caption{(a) Amplitudes of various magnetic structures in the ground state of the Heisenberg Hamiltonian of a ring of alternating sites of $\onehalf$ and $\threehalf$ spins.  In this example we have set the internal Heisenberg coupling $J = 1$, and the external magnetic field $B= 0.1$. 
(b) 
Amplitudes for an alternating sites of spins $\onehalf$ and $\bm{1}$.}
\label{GSLimit}
\end{figure}

The mechanics of the $\mu$-magnons can be described by several phenomena, e.g.,
for a single $\mu$-magnon at site $k$:
\begin{align}
    \ket{\Psi_1} =  A^{\dagger}_{k,k+1} \ket{\text{N\'eel}} ,
\end{align}
a neighboring pair at site $k$ we have:
\begin{align}
   \ket{\Psi_2} =  A^{\dagger}_{k+2,k+3} \ A^{\dagger}_{k,k+1} \ \ket{\text{N\'eel}} ,
\end{align}
for an overlapping pair at site $k$
\begin{align}
    \ket{\Psi_3} = A^{\dagger}_{k+1,k+2}  \ A^{\dagger}_{k,k+1} \  \ket{\text{N\'eel}} ,
\end{align}
and for a split $\mu$-magnon of width $D$:
\begin{align}
    \ket{\Psi_4} = A_{k+1,k+2} \ A^{\dagger}_{k+3+D,k+4+D} \ A^{\dagger}_{k,k+1} \ \ket{\text{N\'eel}} .
\end{align}

In this work we are mostly interested in the quantum correlation properties of $\mu$-magnons, which are created simply through an increase and decrease of the $z$-projection on any odd and even site, respectively. The corresponding creation operator is
\begin{align}
    A_{mn}^\dagger =& 
    \qty( S_+^{\,[m]} \, \delta_{m,\text{odd}} + \bS_-^{\,[m]} \, \delta_{m,\text{even}} ) 
\cdot \qty( \bS_+^{\,[n]} \,  \delta_{n,\text{even}} + S_+^{\,[n]} \, \delta_{n,\text{odd}} ) .
\end{align}
Although magnetic structures can be conveniently studied in terms of $\mu$-magnons, it is essential to emphasize that they should not be regarded as quasiparticles in the strict sense. Instead, they are more analogous to BCS pairs \cite{bardeen1957microscopic}, but going only as far in the comparison as to say that they are composed of two attractive parts rather than being indivisible. While, as we will demonstrate below, they exhibit a form of compactness, their motion is not well-defined due to the alternating spin magnitudes. Additionally, they can partially overlap and are not conserved during the system's evolution.

Let us now discuss \Cref{Amplitudes} in more detail. If the only criterion of choice were the classical energy, we would need to consider a vast number of configurations to include all amplitudes up to $\sim e^{-7}\approx 10^{-3}$, or $\sim e^{-10}\approx 5\times 10^{-5}$. 
The right panel offers a good insight when we introduce $\mu$-magnons. For example, the total number of $\mu$-magnons in a given configuration, as counted by the number of states of spin-$\onehalf$ changed, tends to lower the position of a configuration.
Our goal is to conduct a comprehensive step-by-step analysis of the amplitudes of selected spin configurations.
First, we shall compare results for increasingly larger rings to deduce the thermodynamical limit of $N\rightarrow\infty$. 
Secondly, we will investigate the repetition of certain patterns, which will help us to understand how introducing additional magnetic structures affects the overall amplitude of a configuration.
Finally, we will look into differences due to a physical separation between various structures, thereby describing their mutual dependence.
Unless mentioned otherwise, throughout the rest of this article we will discuss amplitudes relative to the N\'eel state, $\alpha_r(\cdot)$, i.e., for any configuration:
\begin{align}
    \alpha_r(\cdot)=\frac{\alpha(\cdot)}{\alpha(\text{N\'eel})}.
\end{align}
\Cref{GSLimit} demonstrates that the relative amplitudes quickly approach the anticipated values for the thermodynamical limit for $N \gtrsim 14$. Consequently, \Cref{tabela2} presents amplitudes of repeated structures for $N=14$.

\begin{table*}
    \centering
    \begin{tabular}{l||c|c|c|c|c}
         \toprule
         Structure &$k=0$&$k=1$&$k=2$&$k=3$&$k=4$
         \\ \midrule\midrule
$k$ pairs of overlapping ``$\mu$-magnons''&&&&&\\
         $\ket{...,\threehalf,\onehalf,\bm{-\onehalf},\onehalf,\threehalf,\onehalf,\bm{-\onehalf},\onehalf,...}$&1&0.1117&0.012183&0.001335&
          \\ \midrule
         $k$ neighbouring ``$\mu$-magnons''&&&&&\\
         $\ket{...\underbrace{\onehalf,\bm{\onehalf},}_{k\text{ times}}...}$&1&-0.26976&0.084994&-0.027084&0.0086554
          \\ \midrule
         $k$ ``$\mu$-magnons'' separated by a single site&&&&&\\
         $\ket{...,\underbrace{\onehalf\bm{\onehalf},-\onehalf}_{k\text{ times}},...}$&1&-0.26976&0.073473&-0.019589&0.0053334
          \\ \midrule
         $k$ ``$\mu$-magnons'' separated by two sites&&&&&\\
         $\ket{...,\underbrace{\onehalf\bm{\onehalf},-\onehalf,\threehalf,...}_{k\text{ times}},...}$   &1&-0.26976&0.072824&-0.019661 &
         \\ \bottomrule
    \end{tabular}
    \caption{Relative amplitudes $\alpha_r(\cdot)$ of chosen magnetic structures relative to amplitude of the N\'eel state for $N=14$ sites.}
    \label{tabela2}
\end{table*}

\begin{table}
    \centering
    \begin{tabular}{ l||c }
\toprule
         Structure&$\beta'$  
         \\ \midrule\midrule
         $\ket{...,\onehalf,\bm{\onehalf},\onehalf,\bm{\onehalf}...}$&1.1009 \\[0.5em]
         $\ket{...,\onehalf,\bm{\onehalf},-\onehalf,\threehalf,\onehalf,\bm{\onehalf},...}$&1.00025 
         \\ \midrule
         $\ket{...,\onehalf,\bm{\onehalf},\onehalf,\bm{-\onehalf},\onehalf,...}$&1.1184 \\[0.5em]
         $\ket{...,\onehalf,\bm{-\onehalf},\onehalf,\threehalf,-\onehalf,\bm{\onehalf},\onehalf,...}$&1.00046
        \\ \midrule
         $\ket{...,\onehalf,\bm{-\onehalf},\onehalf,\threehalf,\onehalf,\bm{-\onehalf},\onehalf,...}$&0.98197
         \\ \midrule
         $\ket{...,\onehalf,\threehalf,-\onehalf,\bm{\onehalf},\onehalf,\threehalf,-\onehalf,\bm{\onehalf},...}$ & 1.1403  \\[0.5em]
         $\ket{...,\bm{-\onehalf},-\onehalf,\threehalf,\onehalf,\threehalf,\onehalf,\threehalf,\onehalf,\threehalf,-\onehalf,\bm{\onehalf},...}$&0.9474 
          \\ \midrule
         $\ket{...,\onehalf,\threehalf,-\onehalf,\bm{\onehalf},\onehalf,\bm{\onehalf},...}$&1.1426  \\[0.5em]
         $\ket{...,\onehalf,\threehalf,-\onehalf,\bm{\onehalf},-\onehalf,\threehalf,\onehalf,\threehalf,-\onehalf,\bm{\onehalf},...}$&1.0011 \\[0.5em]
         $\ket{...,\onehalf,\bm{\onehalf},-\onehalf,\threehalf,\onehalf,\threehalf,-\onehalf,\bm{\onehalf},..}$&1.0010
         \\ \midrule
         $\ket{...,\onehalf,-\bm{-\onehalf},\onehalf,\bm{\onehalf},-\onehalf,\threehalf,\onehalf}$&1.1761  \\[0.5em]
         $\ket{...,\onehalf,\bm{-\onehalf},\onehalf,\threehalf,-\onehalf,\bm{\onehalf},-\onehalf,\threehalf,\onehalf,...}$&1.0014 \\ \midrule
         $\ket{...,\onehalf,\bm{\onehalf},-\onehalf,\bm{\onehalf},\onehalf,...}$ & 1.0059 \\[0.5em]
         $\ket{...,\bm{\onehalf},\onehalf,\threehalf,\onehalf,\bm{\onehalf}}$&0.9847 
         \\ \midrule
         $\ket{...,\onehalf,\bm{-\onehalf},\onehalf,\threehalf,\onehalf,\bm{\onehalf},...}$&0.9834 
         \\ \midrule
         $\ket{...,\onehalf,\threehalf,-\onehalf,\bm{\onehalf},-\onehalf,\bm{\onehalf},-\onehalf,\threehalf,\onehalf,...}$&1.0130 \\[0.5em]
         $\ket{...,\onehalf,\threehalf,-\onehalf,\bm{\onehalf},-\onehalf,\threehalf,\onehalf,\threehalf,-\onehalf,\bm{\onehalf},...}$&1.0022
         \\ \midrule
         $\ket{...,\onehalf,\threehalf,-\onehalf,\bm{\onehalf},-\onehalf,\bm{\onehalf},\onehalf,...}$&1.0090 \\[0.5em]
         $\ket{...,\bm{\onehalf},-\onehalf,\threehalf,\onehalf,\threehalf,\onehalf,\bm{\onehalf},...}$&0.9719 
         \\ \midrule
         $\ket{...,\onehalf,\bm{-\onehalf},\onehalf,\bm{\onehalf},-\onehalf,\threehalf,\onehalf,...}$&0.9699
         \\ \midrule\\[-1ex]
         spins-$\left(\onehalf,\bm{1}\right)$
         \\ \midrule\midrule
         $\ket{...,\onehalf,\bm{0},\onehalf,\bm{0},...}$&1.2638
         \\ \midrule
         $\ket{...,\onehalf,\bm{0},-\onehalf,\bm{0},\onehalf,...}$&1.018
         \\ \midrule
         $\ket{...,\bm{0},\onehalf,\bm{1},\onehalf,\bm{0},...}$&0.9746
         \\ \bottomrule
    \end{tabular}
    \caption{Ratios of relative amplitudes to sums of products of relative amplitudes of more elementary structures.}
    \label{ampint}
\end{table}

Next, we observe that the relative amplitudes of structures involving multiple $\mu$-magnons could be approximated, with reasonably good accuracy, by a sum of products of relative amplitudes corresponding to their decomposition into more elementary structures. 
For instance, in the case of our 14-site ring:
\begin{align}
    \beta'\left(\ket{...,\onehalf,\bm{\onehalf},\onehalf,\bm{\onehalf},...}\right)=\frac{\alpha_r\left(\ket{...,\onehalf,\bm{\onehalf},\onehalf,\bm{\onehalf},...}\right)}{\alpha_r\left(\ket{...,\onehalf,\bm{\onehalf,...}}\right)^2
    +\alpha_r\left(\ket{...,\onehalf,\bm{\onehalf,...}}\right)
    \alpha_r\left(\ket{...,\onehalf,\threehalf,\onehalf,\bm{\onehalf,...}}\right)}\approx 1.1009 .
\end{align}
This composability provides a useful heuristic for estimating the ground-state amplitude structure. More values of $\beta'$s for various structures are found in \Cref{ampint}. Generally, for separations larger than two sites, the values of $\beta'$ that can be established within our model differ from $1$ by $\lesssim 0.1\%$.

\begin{table*}
    \centering
    \begin{tabular}{l|@{\hspace{5pt}}l||c|c|c|c|c}
         \toprule
         Structure 1 & Structure 2 & $D=0$ & $D=1$ & $D=2$ & $D=3$ & $D=4$
         \\\midrule\midrule    
          $\ket{...,\onehalf,\bm{\onehalf},...}$&$\ket{...,\onehalf,\bm{\onehalf},...}$ , $\ket{...,\bm{\onehalf},\onehalf,...}$&1.1680&1.0097&1.0007&1.0002&1
                \\ \midrule
          $\ket{...,\onehalf,\bm{-\onehalf},\onehalf,...}$&$\ket{...,\onehalf,\bm{\onehalf},...}$ , $\ket{...,\bm{\onehalf},\onehalf,...}$&1.1864&0.98716&1.0010&1&1
                \\ \midrule
          $\ket{...,\onehalf,\bm{-\onehalf},\onehalf,...}$&$\ket{...,\onehalf,\bm{-\onehalf},\onehalf,...}$&&0.9858&&0.9996&
                \\ \midrule
          $\ket{...,\onehalf,\threehalf,-\onehalf,\bm{\onehalf,...}}$&$\ket{...,\onehalf,\bm{\onehalf},...}$ , $\ket{...,\bm{\onehalf},\onehalf,...}^{(a)}$&1.2917&1.0171&1.0036&1.0004&1.0013
                \\ \midrule
          $\ket{...,\onehalf,\threehalf,-\onehalf,\bm{\onehalf,...}}$&$\ket{...,\onehalf,\bm{\onehalf},...}$ , $\ket{...,\bm{\onehalf},\onehalf,...}^{(b)}$&1.2847&0.9797&1.0035&0.9993&1.0013
                \\ \midrule
          $\ket{...,\onehalf,\threehalf,-\onehalf,\bm{\onehalf,...}}$ , $\ket{...,\bm{\onehalf},-\onehalf,\threehalf,\onehalf,...}$&$\ket{...,\onehalf,\bm{-\onehalf},\onehalf,...}$&1.3274&0.9782&1.0035&1&
              \\ \midrule
          $\ket{...,\onehalf,\threehalf,-\onehalf,\bm{\onehalf,...}}$ , $\ket{...,\bm{\onehalf},-\onehalf,\threehalf,\onehalf,...}$&$\ket{...,\onehalf,\threehalf,-\onehalf,\bm{\onehalf,...}}$ , $\ket{...,\bm{\onehalf},-\onehalf,\threehalf,\onehalf,...}$&1.9055&1.0303&1.1330&0.9993&\\
          &&&(spin-${\onehalf})$&&&\\
          &&&0.9636&&&\\
          &&&(spin-$\threehalf)$
          \\ \bottomrule
    \end{tabular}
    \caption{Values of $\beta$'s f, various pairs of magnetic structures in function of separations $D$ between them. (a) at the side of $\ket{...,\onehalf,\bm{-\onehalf},...}$. (b) at the side of $\ket{...,\threehalf,\onehalf...}$.}
    \label{ampint1}
\end{table*}

It is also interesting to discuss ratios $\beta(\text{str}_1,\text{str}_2,D)$, where $\text{str}_1$, $\text{str}_2$ are two magnetic structures and $D$ is the separation between them. They are given by 
\begin{align}
\beta(\text{str}_1,\text{str}_2,D)=\frac{\alpha_r(\text{str}_1,\text{str}_2,D)}{\alpha_r(\text{str}_1)\alpha_r(\text{str}_2)} ,
\end{align}
and numerically listed in \Cref{ampint1}. As the amplitude of a split $\mu$-magnon quickly decays with the separation, these corrections seem significant only at short distances.

Magnetic structures that are far apart from each other can be seen as independent, i.e., a relative amplitude of well-separated structures is a product of their individual relative amplitudes.
This can be related to the amplitudes of $\mu$-magnons in the function of separation $D$, as listed in \Cref{split1}.
While \Cref{tabela2} demonstrates that a multiplication of structures on the N\'eel sea approximately decreases amplitude by an appropriate factor. Thus they could seen as almost independent of each other. \Cref{ampint1} investigates how these factors change in the function of separation between the structure.

Lastly, we address the open-end problem by comparing its ground state properties with that of the ring.
Considering a total of 14-sites, the relative amplitudes of individual $\mu$-magnons, split  or non-split are plotted in \Cref{OpenEnds}. 
Significant effects arise only when a spin-$\onehalf$ is located at the edge.
A study of the open-end configuration allows us to investigate the decay of the amplitudes of split $\mu$-magnons as a function of their separation; this decay follows a sub-exponential trend. 
A fit to the average behavior, excluding configurations involving the edge spin-$\onehalf$, reads:
\begin{align}
\alpha_r(\text{split} \ \mu\text{-magnon}) \approx 
 - 0.27295 \times
 \exp(-1.55089D +   0.07923D^2 )  , \quad : D \leq 10.
\label{eq_relamp_approx}
\end{align}
While it would be impractical to present all possible cases, the provided examples correctly capture the underlying universal patterns and dependencies.

\begin{table*}
    \centering
    \begin{tabular}{c||c|c|c|c}
\toprule 
         Split $\mu$-magnon & $D=0$ & $D=2$ & $D=4$ & $D=8$ \\
\midrule\midrule
         $\alpha_r$ \, spins-($\onehalf,\threehalf$) & $-2.697 \times 10^{-1}$ & $-1.643  \times 10^{-2}$ & $-2.145 \times 10^{-3}$ & $-6.719 \times 10^{-4}$ 
         \\ \midrule
         $\alpha_r$ \, spins-($\onehalf,\bm{1}$) & $-3.211 \times 10^{-1}$ & $-2.651  \times 10^{-2}$ & $-4.8878 \times 10^{-3}$ \\          
\bottomrule
    \end{tabular}
        \caption{Relative amplitudes of a split $\mu$-magnon with gap width $D$.}
    \label{split1}
\end{table*}

\section{Truncation and entanglement of reduced states}
\label{trunc}

With an improved understanding of the mechanisms underlying the ground state of the spin-$\qty(\onehalf, \threehalf)$ Heisenberg chain, we are now in a position to examine reduced states and their correlations. This analysis aims to enable a meaningful truncation of the Hilbert space to facilitate the study of even larger systems, also in combination with techniques such as DMRG.

\begin{figure}[!b]
    \centering
    \includegraphics[width=0.55\linewidth]{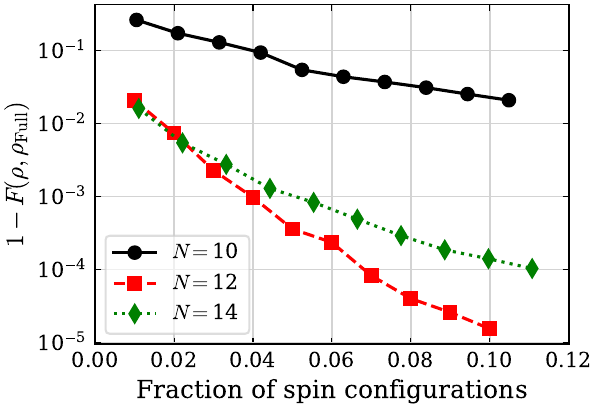}
    \caption{Infidelity of a truncated reduced states of four consecutive sites in function of the fraction of included spin-$\left(\frac{1}{2},\mathbf{\frac{3}{2}}\right)$ configurations for $N=10$ (blue), 12 (orange) and 14 (green).}
    \label{Infidelity}
\end{figure}

Let us now consider the infidelity, 
between the actual reduced state of four neighboring sites $\rho_{\text{Full}}$ and the truncated state $\rho$, which includes a fraction of the most relevant spin configurations:
\begin{align}
    1-F(\rho,\rho_\text{Full})=1-\left(\text{tr}\left(\sqrt{\sqrt{\rho_{\text{Full}}\rho}\sqrt{\rho_{\text{Full}}}}\right)\right)^2 .
\end{align}
For this comparison, we have considered $N$-site ($N=10,12,14$) spin-$\left(\frac{1}{2},\mathbf{\frac{3}{2}}\right)$ rings. Figure \ref{Infidelity} shows while the infidelity significantly increases for $N=12,14$ as compared to 10, more configurations can be required for $N=14$ to obtain the same infidelity as for $N=12$. Still, this results suggest that it may suffice in general to take about 5\% of all configurations for satisfactory fidelities.

\begin{figure}
\centering

\subfigure[]
{ \includegraphics[width=0.55\linewidth]{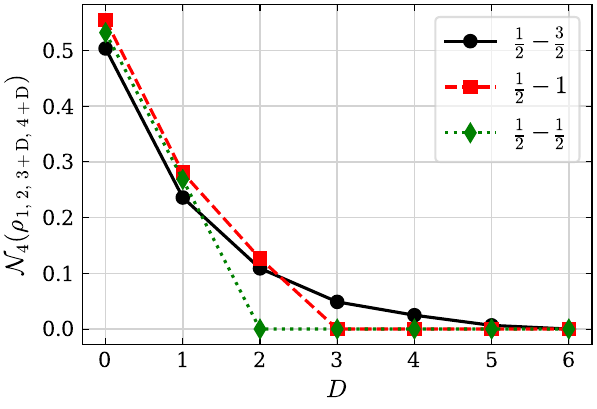} }

\subfigure[]
{ \includegraphics[width=0.55\linewidth]{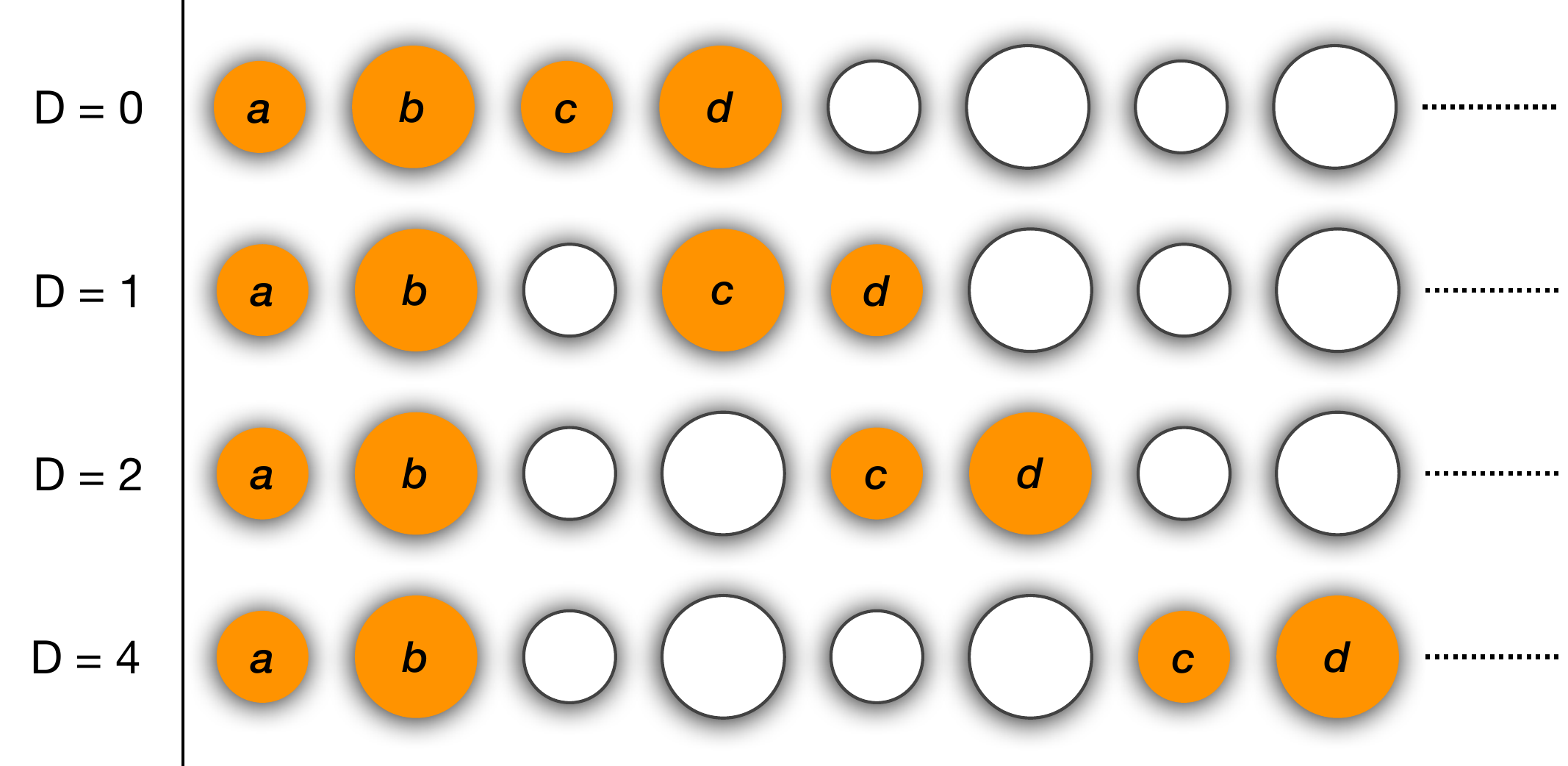} }

\caption{(a)
Four-partite negativity between two separated spin pairs in the ground state of a very-long closed chain of alternating spins $\onehalf$ - $\threehalf$, $\onehalf$ - $\bm{1}$, and $\onehalf$ - $\onehalf$. 
The reduced density matrix of the four sites, two spin pairs, is denoted by $\rho_{abcd}$. For simplicity the first spin pair is chosen at the beginning of the chain: $(a,b) = (1,2)$. The second spin pair is separated by a distance $D$: (c,d) = $(3+D,4+D)$, as illustrated in the bottom panel (b). 
For this example we have set the internal Heisenberg coupling $J = 1$ and the external magnetic field $B= 0.1$. }

\label{4partite}
\end{figure}

The above calculations suggest that it is possible to obtain highly precise results with a relatively strong truncation of the ground state. Let us consider two pairs of neighboring sites. We will verify the presence of genuine four-site entanglement. The figure of our interest will be four-partite negativity, which is the straight-forward generalization of the tripartite negativity introduced by Sabin {\em et al.} \cite{sabin2008classification}:
\begin{align}
\mc{N}_4(\rho) 
=& \Bigg( \prod_{\mc{S}} \log_2 \bigg( 1+2\sum_{\lambda_k^{(\mc{S})}<0}\lambda_k^{(\mc{S})} \bigg) \Bigg)^{1/7} ,
\quad\quad   : 
\mc{S} \in \qty{a,b,c,d,ab,ac,ad}   .
\end{align}
where $\rho = \rho_{abcd}$
is the four-site reduced density matrix, and $\lambda^{(\mc{S})}$ are the eigenvalues of $\rho^{\Gamma(\mc{S})}$, which is simply the partial transposition of $\rho$ with respect to $\mc{S}$. While the four-partite negativity is not a sufficient criterion of genuine four-site entanglement, the nonvanishing of its three-body counterpart certifies the ability to distill GHZ states \cite{sabin2008classification}.
The results are presented in \Cref{4partite}
show that, while for the ground state of the Heisenberg chain of spins-$\onehalf$ the four-partite negativity is restricted to nearest neighbors, in ferrimagnetic systems it is present even at large separations.

Figure \ref{4partite} shows that genuine four-site entanglement, at least in the form detected by $\mc{N}_4$, has a long-range nature, compared to the spin-$\onehalf$ chain. In other words, if the system has a phase transition between effective spins-$\onehalf$ and spins-$\threehalf$, this phase transition can be also seen via the presence of long-range quantum correlations.

\begin{figure}
    \centering
    \includegraphics[width=0.5\linewidth]{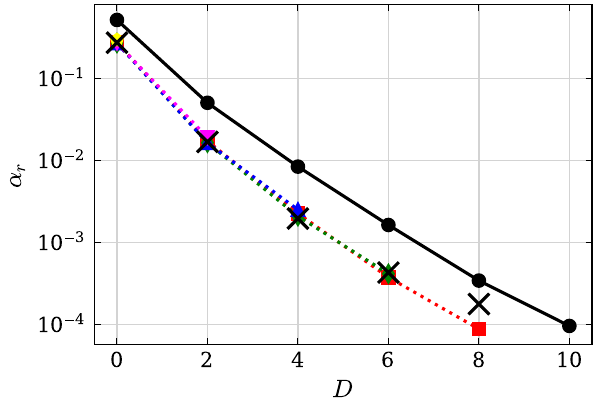}
    \caption{Relative amplitudes of split $\mu$-magnons of separation $D$ in the ground state of 12-site open-end chain. Solid black: spin-$\onehalf$ at the end of the chain has value $\onehalf$. Dashed lines in other colors: all other cases, i.e., configurations with value $-\onehalf$, with the scattered black crosses showing their analytical approximations as calculated through \Cref{eq_relamp_approx}.}
    \label{OpenEnds}
\end{figure}

Lastly, we shall consider the effects of open ends. For this reason, we have computed the ground state for {\em open-end} chain of 14 sites to compare it to the ring solution. In particular, we are interested in relative amplitudes of a single $\mu$-magnon, split or non-split. Regardless of the gap between the two altered spins, $\alpha_r$ is almost equal to the ring case, except for when the flipped spin-$\onehalf$ is at the first site of the chain. We also take this opportunity to study the decay of the amplitudes of the $\mu$-magnons. As can be seen in \Cref{OpenEnds}, the differences between amplitudes of $\mu$-magnons are almost independent of the distance from the end, except for the case in which the $\mu$-magnon involves the spin-$\onehalf$. 

\begin{figure}
\centering
    
\includegraphics[width=0.6\linewidth]{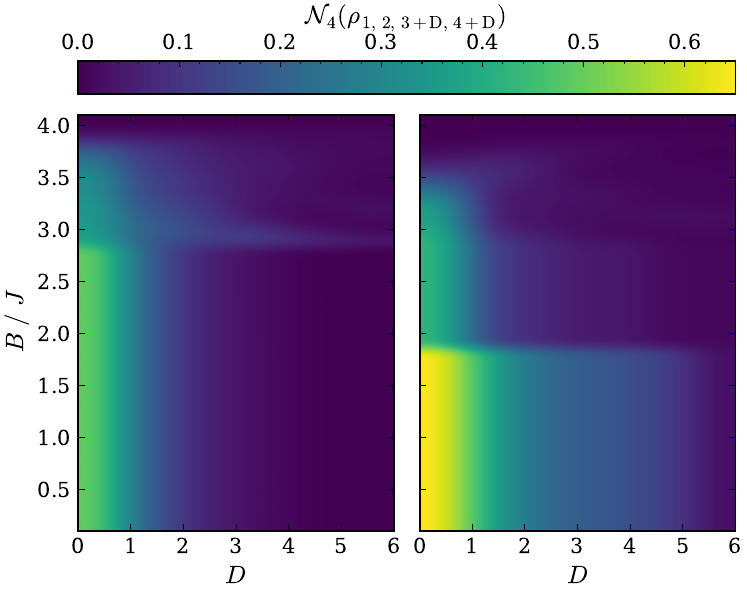}
    
\caption{Ground state in closed (left) and open (right) chain of very-long chain of alternating spins $\onehalf-\threehalf$, as calculated through the DMRG method. $\mc{N}_4$ denotes the four-partite logarithmic negativity, $D$ is the separation of spin pairs, and $B/J$ is the ratio of the applied external magnetic field to the internal Heisenberg coupling.}
\label{FourNegDensityPlot1}
\end{figure}

Using DMRG,  we confirm for the weak magnetic field that 4-site entanglement is higher when the end spin-$\onehalf$ is involved (\Cref{FourNegDensityPlot1}). A stronger magnetic field partially overcomes the Heisenberg interaction and changes the total angular momentum of the ground state. Interestingly, this happens for different values of $B$ for a ring and a chain, and with different patterns.

\section{Extension to longer chains}

The ultimate goal of this analysis is to study presence of entanglement in the thermodynamical limit. While periodic boundary conditions introduce higher symmetry, one needs to consider larger systems to eliminate any effects of boundaries. The data presented above justifies the following observations, all within some approximation. First, amplitudes of analogous spin configurations, relative to the amplitude of the N\'eel state stabilize in the thermodynamical limit, with a ring of 14 sites being reasonably close to the asymptotic values. Secondly, introducing a new magnetic structure to a spin configuration decreases the relative amplitude by a factor corresponding to this factor. Thirdly, the relative amplitude of a split $\mu$-magnon decreases rapidly with the gap. Finally, there are additional factors depending on the separation between two structures, but it rapidly approaches 1 as the separation grows.
\begin{align}
\label{alpha1}
    \alpha_r(\text{str}_1,\text{str}_2,D) &= \tilde{\alpha}_r(\text{str}_1) \ \tilde{\alpha}_r(\text{str}_2) \ \tilde{\beta}'(\text{str}_1,\text{str}_2,D), 
    \nonumber\\[0.5em]
    \alpha_r(\text{str}_1,D_{1,2},\text{str}_{2},...,\text{str}_N,D_{N,1}) &= \prod_{i=1}^N\tilde{\alpha}_r(\text{str}_i)  \prod_{j}^N\tilde{\beta}'(\text{str}_j,\text{str}_{j+1},D_{j,j+1}) ,
\end{align}
where $D_{j,j+1}$ is a gap between structures $\text{str}_j$, $\text{str}_{j+1}$, $n+1\equiv 1$, and $\tilde{\alpha}_r$, $\tilde{\beta}'$ are taken from a dictionary, e.g., the solution for $14$ sites. Factors related to boundary conditions are already included in $\alpha_{r}$. If a single structure is longer than $7$ sites, its $\tilde{\alpha}$ can be taken as 0, as it does not appear in the dictionary. If, however, the separation between two structures is so large that the configuration cannot be interpreted within the dictionary, the respective $\tilde{\beta}'$ is taken as 1.

Considering larger systems will inevitably lead to an exponential number of configurations to be considered, quickly saturating the computational capabilities of any machine. Therefore, like in the case of renormalization groups, one needs to find efficient ways to leave only the most relevant few. In Section \ref{trunc} we have shown that the essential information is contained in a fraction of the most probable configurations. First, we need to choose a reasonable threshold,  below which relative amplitudes (in moduli) can be rejected. The case of $N=14$ suggests the cut-off at $10^{-3}$, leaving 2563 configuration, about $10\%$ of the total.

Having this threshold, we can implement some criteria to reject certain configurations a priori. For example, a configuration should host at most six $\mu$-magnon, provided that they create three overlapping pairs. Likewise, an acceptable configuration should feature not more than two spins-$\threehalf$ with value $\bm{-\threehalf}$. Notice that while the amplitude threshold may not be universal for higher $N$, limiting the number of $\mu$-magnons changes the complexity of the problem from exponential to polynomial in terms of configurations to be considered. The next step is to check if any leftover configuration contains a sub-configuration from the dictionary, which already has $|\alpha_r|$ below the threshold, as introducing more structures can only decrease when more structures are introduced.

Finally, we proceed with estimating amplitudes for the remaining configuration. An algorithm would scan the states of subsequent spins. A structure begins when the cumulative magnetization 
 diverges from this of the N\'eel state:
 \begin{align}
      m_j = \sum_{k=1}^j \langle S_{z}^{[k]}\rangle-\sum_{k=1}^j\langle S_{z}^{[k]}\rangle_{\ket{\text{N\'eel}}} \neq 0 ,
 \end{align}
where 
 \begin{align}
    \sum_{k=1}^j \langle S_{z}^{[k]}\rangle_{\ket{\text{N\'eel}}} =
     \begin{cases}
        \frac{j}{2} - 1, & j \in \text{odd}, \\[1ex]
        \frac{j}{2}, & j \in \text{even}.
     \end{cases}
 \end{align}
The structure ends when we reach $m_{j}=0$. In the same way we find gaps between structures and compute the respective relative amplitudes according to Eq.~\eqref{alpha1}. A configuration is scanned both ways, and the maximum of the two amplitude estimates is chosen. The remaining amplitudes constitute an approximate ground state.

\begin{figure}
    \centering
    \includegraphics[width=0.55\linewidth]{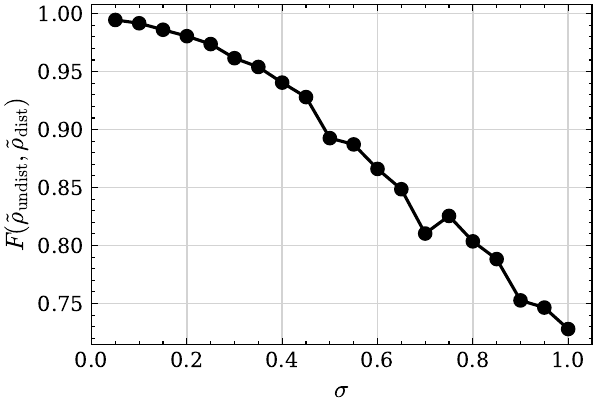}
    \caption{Fidelity between a reduced state of four consecutive sites of a 12-site ring with and without a random change. Each point is averaged over 40 instances.}
    \label{distorted}
\end{figure}

We also include the calculations for the fidelity of between a distorted reduced state $\tilde{\rho}_{\text{dist}}$ of four neighboring sites of a 12-site  ring to undistorted state $\tilde{\rho}_{\text{undist}}$ as a function of $\sigma$, which we defined as the standard deviation of a random variable ($x$) that follows a normal distribution. Then, in $\tilde{\rho}_{\text{dist}}$ each $\alpha$ is multiplied by $\text{exp}(x)$. This is relevant in the sense that we artificially take $\beta$'s equal to 1 for larger separation.
The corresponding results are presented in \Cref{distorted}, and they show that the model is not very sensitive for such changes.

\section{Summary and Conclusions}

While ferrimagnets are crucial in several existing technologies, their microscopic behavior is still under-explored.
We have investigated the exact numerical ground states of antiferromagnetic rings of spins of alternating magnitudes $\onehalf$ and $\threehalf$. 

Based on the exact numerical solution for up to 14 sites, we introduce a novel description of $\mu$-magnons, an elementary deviation from the N\'eel pattern. 
We found that the amplitudes of spin configurations are very well approximated in terms of the amplitudes of more basic magnetic structures, along with factors dependent on their spatial separation. Furthermore, it has been demonstrated that the amplitudes of at least specific magnetic structures, relative to the amplitude of the N\'eel sea, stabilize rapidly for $ N\rightarrow\infty$.
Subsequently, we have shown that for a 14-site ring, it is sufficient to consider relatively few spin configurations for a faithful reconstruction of the reduced states of two and four sites.

Therefore, we conclude that the interesting quantum properties of larger systems of this kind can be captured by configurations involving only a few $\mu$-magnons. 
Regarding quantum correlations, two-body entanglement is not observed beyond neighboring sites, but we observe long-range multipartite entanglement between pairs of spatially separated regions. 
Introducing open boundary conditions further extends this range, which is in contrast with the case of a spin-$\onehalf$ ring, where this form of entanglement vanishes beyond a distance of 2 sites. The results, obtained by truncating the reduced density matrices to a relatively small number of the most important $\mu$-magnon configurations, agree with DMRG calculations. 
While the amount of entanglement does not change drastically with the magnitude of the larger spins, it decays exponentially with the separation between pairs of sites and vanishes at larger distances for systems with larger spin magnitudes.

The presented $\mu$-magnon description does not go beyond the scope of the results obtained with DMRG, but it is valuable on many levels due to an intuitive understanding, such as in the case of entanglement at the boundary. 
It also provides a good starting point for variational studies of entanglement between very distant sites, where the DMRG treatment has an exponentially large computational cost.
Additionally, this approach may find interesting applications in studying ground states of two-dimensional lattices with diverse topologies.

\begin{acknowledgments}
   This work is supported by NCN SONATA-BIS grant No: 2017/26/E/ST2/01008. IHNN acknowledges partial support by the Foundation for Polish Science (IRAP project, ICTQT, Contract No. MAB/2018/5, co-financed by EU within Smart Growth Operational Programme).
\end{acknowledgments}


%

\clearpage
\appendix

\section{Most significant configurations for a $N=14$ ring}

\begin{figure*}[!h]
    \centering
    \includegraphics[width=0.66\linewidth]{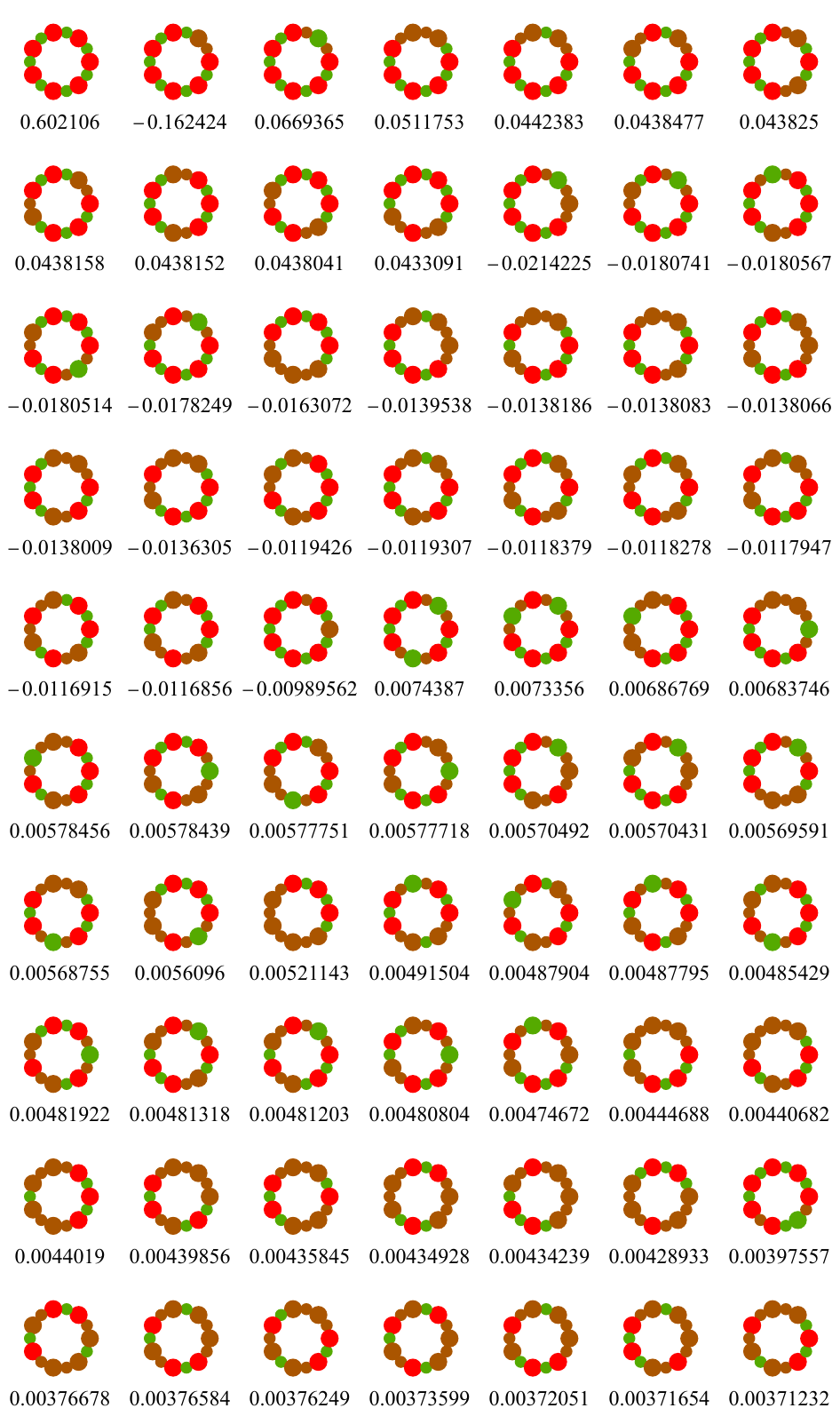}
    \caption{70 examples of most probable spin configurations in the ground state (each representing its own orbit of translational and reflective symmetries) for $N=14$, together with respective absolute amplitudes. The graphical coding is the same as in \Cref{structure}(a).}
    \label{config}
\end{figure*}

\end{document}